\documentstyle[epsfig]{aipproc}

\begin{document}
\title{Macroscopically large antimatter regions\\ in the baryon asymmetric universe}

\author{ Alexander Sakharov$^*$, Maxim Khlopov${^\dagger}$$^{\ddagger}$ and Sergei Rubin$^{\dagger}$$^{\ddagger}$
}
\address{$^*$Labor f\"ur H\"ochenergiephysik, ETH-H\"onggerberg, 
HPK--Geb\"aude, CH--8093 Z\"urich\\
$^{\dagger}$ Center for CosmoParticle Physics "Cosmion", 4 Miusskaya pl., 125047 Moscow, Russia\\ 
$^{\ddagger}$ Moscow Engineering Physics Institute, Kashirskoe shosse 31, 115409 Moscow, Russia}

\maketitle

\begin{abstract}
The existence of macroscopic regions with antibaryon excess in
the matter -- dominated Universe is a possible
consequence of the evolution of baryon charged, pseudo -- Nambu --
Goldstone field with lepton number violating couplings. Such regions can
survive the annihilation with surrounding matter only in the case if their
sizes exceeds the critical surviving size. The evolution of  survived
antimatter -- regions with high original antibaryon density inside results
in the formation of globular clusters, which is made out from antimatter
stars. The origin of  antimatter regions in the chosen scenario is
accompanied the formation of closed domain walls, which can collapse into
massive black holes deposed inside the high density antimatter regions.
This fact can give us an additional hint, that an anti -- stars globular
cluster could be one of the collapsed -- core star clusters, which
populate our galaxy. \end{abstract}

\section*{Introduction}

Since a long time the generally accepted motivation for baryon asymmetric
Universe were the direct observations, which claim to exclude the
macroscopic amount of antimatter within the distance up to 20Mpc from the
solar system \cite{exl}. Moreover, if larger than 20Mpc regions of matter
and antimatter coexist, then it would be impossible to keep them out of
the close contact during an early time, because the uniformity of CMBR
excludes the existence of any significant voids. 
The 
annihilation, which would take place at the border between matter and
antimatter region, during the period $1100>z>20$, would disturb the 
diffuse $\gamma$ -- ray background
\cite{crg}, if the size of matter or antimatter regions does not exceed
$10^3$Mpc. Thus the baryon symmetric Universe is practically excluded.
However, such arguments cannot exclude the case when the Universe is
composed almost entirely of matter with relatively small insertions of
primordial antimatter. Thus we could expect the existence of
macroscopically large antimatter regions in the baryon asymmetric universe
as a whole.  We call such a region the local antimatter area (LAA). 

Any
primordial LAA having initial size up to $\simeq 1$pc or more at the end
of radiation dominated (RD) stage is survived the boundary annihilation
with surrounding matter until the contemporary epoch \cite{we} and in the
case of successive homogeneous expansion has the critical surviving size
$l_c\simeq 1kpc$ or more. The smaller LAA's will be eaten up by the
annihilation. This fact makes problematically to apply any model with
usual thermal phase transition to modulate primordial matter antimatter
distribution over the size exceeding $\l_c$ \cite{dolg,zil}. We could
think about a possible inflational blow upping of the correlation length
of a usual phase transition \cite{dolsil}, but one should also take care
about some unwanted topological defects, which could accompany phase
transitions and significantly contribute to the energy density of the 
universe. Mostly, to
get rid from unwanted topological defects some mechanisms of symmetry
restoration should be invoked \cite{dolsil}. Here we present the issue for
inhomogeneous baryogenesis \cite{zil}, which is free from the difficulties
connected with usual phase transition approach and able to generate a
considerable number of above -- critical LAAs, what makes reasonable to
discuss the existence of primordial LAA in our galactic volume.

\section*{Formation and evolution of LAA's}

\paragraph*{The Formation scenario.} Our antimatter generation scenario \cite{zil} is  based on the
spontaneous baryogenesis mechanism \cite{sb}, which implies the existence
of   complex scalar field $\chi =(f/\sqrt{2})\exp{(\theta )}$ carrying
baryonic charge with explicitly broken $U(1)$ symmetry. The explicit
breakdown of $U(1)$ symmetry is coming from the phase depended term,
which tilts the bottom of the Nambu -- Goldstone (NG) potential. We
suppose \cite{zil} that the radial mass $m_{\chi}$ of field $\chi$ is
larger then the Hubble constant $H$ during inflation, while for the
angular mass of $\chi$ just the opposite condition $m_{\theta}\ll H$ is
satisfied at that period. It makes sure that $U(1)$ symmetry is
already broken spontaneously at the beginning of inflation, but the
background vacuum energy is still so high, that the tilt of the potential
is vanished. This implies that the phase $\theta$ behaves as ordinary
massless NG boson and the radius of NG potential is firmly established by
the scale $f$ of spontaneous $U(1)$ symmetry breaking. Owing to quantum
fluctuations of effectively massless angular component $\theta$ at the de
Sitter background \cite{linde} the phase $\theta$ is varied in different
regions of the Universe. Actually, such fluctuations can be interpreted as
the one -- dimensional Brownian motion \cite{linde} along the circle
valley corresponding to the bottom of the NG potential. When the vacuum
energy decreases the tilt of potential becomes topical, and pseudo NG
(PNG) field starts oscillate. Let us assume that the phase value $\theta
=0$ corresponds to South Pole of  NG field   circle valley, and $\theta
=\pi$ corresponds to the opposite pole. The positive gradient of phase in
this picture is routed as counterclockwise direction, and the dish of PNG 
potential  would locate at the South Pole of circle. The possible
interaction of field $\chi$ that violates the lepton number can have such
a structure \cite{zil,dolgmain}, that as the $\theta$ rolls down in
clockwise direction during the first oscillation, it preferentially
creates baryons over antibaryons, while the opposite is true as it rolls
down in the opposite direction during the first oscillation. Thus to have
the globally baryon dominated Universe one must have the  phase sited in the
range $[\pi ,0 ]$, just at the beginning of inflation  (when the size of
the modern Universe crosses the horizon).  Then subsequent quantum
fluctuations  move the phase to some points $\bar\theta_i$ at the range
$[0,\pi ]$ causing the antibaryon excess production. If it takes place not
later then after 15 e -- folds from the beginning of inflation
\cite{zil}, the size of LAA's will exceed the critical surviving size
$l_c$. Let set the phase at the point $\theta_{60}$ in the range $[\pi
,0 ]$, where for simplicity we suggest that the total number of
inflational e -- folds is 60. The phase makes Brownian step $\delta\theta 
=H/(2\pi f)$  at each e--fold. Because the typical wavelength of the 
fluctuation $\delta\theta$   generated during such timescale is equal to
$H^{-1}$, the whole domain  $H^{-1}$,   containing $\theta_{60}$, after
one e--fold effectively becomes divided into  $e^3$  separate, causal
disconnected domains of radius $H^{-1}$. Each domain contains almost
homogeneous  phase value  $\theta_{60-1}=\theta_{60}\pm\delta\theta$. In
half of these  domains the phase evolves towards $\pi$ (the North Pole)
and in the other  domains  it moves  towards zero (the South Pole). To
have LAA's with appropriate  sizes to avoid full annihilation one  should 
require that the phase value crosses  $\pi$ or zero  not later then after
$15$ steps. The numerical calculations \cite{zil} of the domain size
distribution filled with appropriate phase values $\bar\theta_i$ show that a
volume box corresponding to each galaxy can contains 1--10 above -- critical 
regions with
appropriate  phase $\bar\theta_i$ at the condition that the fraction of
the universe containing $\bar\theta_i$ is many orders of magnitude less
then 1 \cite{zil}. The last conclusion makes sure that the universe will
become baryon asymmetric as a whole. At the some moment after the end of
inflation deeply at the Friedman epoch the condition $m_{\theta}\ll H$ is
violated and the oscillations of $\theta$ around the minima of PNG
potential are started. Then the stored energy density $\rho_{\theta}\simeq
\theta^2m_{\theta}^2f^2$ will convert into baryons
and antibaryons. All domains where the phase starts to oscillate from the
values $\bar\theta_i$ will contain antimatter. The density of antimatter
depends on the initial value $\bar\theta_i$ and can be different in the
different domains \cite{zil}. The average number density of surrounding
matter should be normalised on the observable one $n_B/s\simeq 3\cdot
10^{-10}$. This normalisation sets the condition $f/m_{\theta}\ge 10^{10}$
for the PNG potential \cite{zil}.
   
\paragraph*{Anti -- star globular cluster formation.}
At 
the condition $f\ge H\simeq 10^{13}$GeV \cite{zil} we can have a considerable number of high density above --
critical LAA's that makes sense to discuss the possible evolution of such
a LAA in our galaxy. It is well known \cite{glob} that clouds, which have
temperature near $10^4$K and densities several ten times that of the
surrounding hot gas, are gravitationally unstable if their masses are of
the order of $10^5M_{\odot}-10^6M_{\odot}$. These objects are identified as the
progenitor of globular cluster (GC) and reflect the Jeans mass at the
recombination epoch. From the other side the typical size of that mass is
close to the $l_c$ at the end of RD epoch. Thus if the primordial
antibaryon density inside  a LAA was one order of magnitude higher then
surrounding matter density, that LAA can evolve into antimatter GC
\cite{khl}. Moreover, to imprint a characteristic Jeans mass the proto --
GC must cool slowly \cite{glob} after the recombination, so the heating of
dense antimatter might be supported by annihilation with surrounding
matter. Thereby GC at the large galactocentric distance is the ideal
astrophysical objects which could be made out of antimatter, because GC's
are the oldest galactic system to form in the universe, and contain stars
of the first population. The existence of one of such anti -- star GC with the mass $10^3M_{\odot}-10^5M_{\odot}$ will not
disturb observable $\gamma$ -- ray background \cite{khl}, but the expected
fluxes of $\overline{^4He}$ and $\overline{^3He}$ from such an antimatter
 object \cite{bgk} are only factor two below the limit of AMS--01 (STS--91) 
experiment \cite{ams} and definitely accessible for the sensitivity of
coming up AMS--02 experiment. 

\paragraph*{Topological defects and black holes (BH).}
The angular term of $\chi$ potential
$m_{\theta}^2f^2(1-\cos\theta)$, which breaks $U(1)$ symmetry explicitly
has a number of discrete degenerate minima \cite{bh}. The equation of
motion with such a potential admits a kink -- like, domain wall (DW)
solution, which interpolates between two adjacent vacua, for example
between $\theta =0$ and $\theta =2\pi$. From the other side our scenario
\cite{zil} deals with the situation when at the beginning of inflation the
universe contains the uniform $\theta$ in the interval $[\pi ,0]$ and
hence the final vacuum state of baryon asymmetric part is $\theta =2\pi$.
On the contrary, there will be the island with $\theta$ in the range
$[0,\pi ]$ where the phase came trough the North Pole due to the
fluctuations. The phase inside that islands will produce preferentially
antimatter and come to the vacuum state $\theta =0$. Thus both states
$\theta =2\pi$ and $\theta =0$ are separated by closed DW's. The collapse
of such a DW is unavoidable \cite{bh}, and DW's which are generated
before 20 inflation e -- folds will form BH's. The
density profile, concentration and observable central cusp in the stars
velocity dispersion of the collapsed -- core clusters GC  NGC 7078 (M15) 
consist with the hypothesis
of the massive $\approx 10^3M_{\odot}$ central BH existence \cite{glob}.
This mass corresponds to the 33 inflation e -- folds. It means that the
DW was originally already encompassed the size $\l_c$ giving rise the
central BH formation, which could induce the collapsed -- core properties of M15. 

\paragraph*{Acknowledgments.} 
S.R. and M.K. work at the project 
``Cosmoparticle Physics'' and  acknowledge support from Cosmion -- ETHZ collaboration.


\begin{references}
\bibitem{exl}Steigman, G. A., {\it Ann.\ Rev.\ Astron.\  Astrophys.}\ {\bf
14}, 339 (1976).
\bibitem{crg}Cohen, A. G., De Rujula, A., and Glashow, S. L., {\it Astrophys.\ J.} {\bf 495}, 539 (1998).
\bibitem{we}Khlopov, M. Yu., et al,   
{\it Astropart.\ Phys.} {\bf 12}, 367 (2000).
\bibitem{dolg}Dolgov, A. D., hep -- ph/9605280; {\it Phys.\ Rep.} {\bf 222}, 309 (1992).
\bibitem{zil}Khlopov, M. Yu., Rubin, S. G., and Sakharov, A. S., {\it Phys.\ Rev.} {\bf D62}, 083505 (2000).
\bibitem{dolsil}Kuzmin, V., Tkachev, I., and Shaposhnikov, M., {\it Phys.\ Lett.} {\bf 105B}, 167 (1981).
\bibitem{sb}Cohen, A. G., and Kaplan, D. B., {\it Phys.\ Lett.} {\bf B199}, 251
(1987); {\it Nucl.\ Phys.} {\bf B308}, 913 (1988).
\bibitem{dolgmain}Dolgov, A. D.,  et al, {\it Phys.\ Rev.} {\bf D56}, 6155 (1997). 
\bibitem{linde}Linde, A., {\it Particle\ Physics\ and\ Inflationary\  Cosmology}: Harwood, 1990.
\bibitem{glob}Meylan, G., and Heggie, D. C., {\it Astron.\ Astrophys.\ Rev.} {\bf 8}, 1
(1997).
\bibitem{khl}Khlopov. M. Yu., {\it Gravitation \& Cosmology} {\bf 4}, 1 (1998).
\bibitem{bgk}Belotsky, K. M., et al, 
{\it Phys.\ Atom.\ Nucl.} {\bf 63}, 233 (2000); {\it astro--ph/9807027}.
\bibitem{ams}AMS Collaboration, Alcaraz, J., et al, {\it Phys.\ Lett.} {\bf 461B}, 387 (1999).
\bibitem{bh}Rubin, S. G., Khlopov, M. Yu., and Sakharov A, S., {\it hep-ph/0005271}.
\end{references}
\end{document}